\documentclass{emulateapj}
\usepackage{apjfonts}
\usepackage{graphicx}

\shorttitle{Cygnus Loop}
\shortauthors{Salvesen et al.}

\begin{document}

\title{Shock Speed, Cosmic Ray Pressure, and Gas Temperature in
the Cygnus Loop}

\author{Greg~Salvesen\altaffilmark{1,2}, John~C.~Raymond\altaffilmark{1}, 
Richard~J.~Edgar\altaffilmark{1}}

\altaffiltext{1}{Harvard-Smithsonian Center for Astrophysics, 60
Garden Street, Cambridge, MA 02138; salvesen@head.cfa.harvard.edu,
jraymond@cfa.harvard.edu, edgar@head.cfa.harvard.edu}
\altaffiltext{2}{Department of Astronomy, The University of Michigan,
500 Church Street, Ann Arbor MI, 48109-1042; salvesen@umich.edu}

%\authoremail{salvesen@umich.edu}

\label{firstpage}

\begin{abstract}
Upper limits on the shock speeds in supernova remnants
can be combined with post-shock
temperatures to obtain upper limits on the ratio of cosmic ray to gas
pressure ($P_{CR} / P_G$) behind the shocks.  We constrain shock
speeds from proper motions and distance estimates, and we 
derive temperatures from X-ray spectra.
The shock waves are observed as faint H$\alpha$
filaments stretching around the Cygnus Loop supernova remnant in two
epochs of the Palomar Observatory Sky Survey (POSS) separated by 39.1
years.  We measured proper motions of 18
non-radiative filaments and derived shock velocity limits based on 
a limit to the Cygnus Loop distance of 576 $\pm$ 61 pc given by Blair
et al. for a background star.  The PSPC instrument
on-board {\it ROSAT} observed the X-ray emission of the post-shock gas
along the perimeter of the Cygnus Loop, and we measure post-shock
electron temperature from spectral fits. Proper motions range from 2\farcs 7 to
5\farcs 4 over the POSS epochs and post-shock temperatures range from $kT
\sim 100-200 \rm~eV$.  Our
analysis suggests a cosmic ray to post-shock gas pressure
consistent with zero, and in some positions $P_{CR}$ is formally smaller
than zero.  We conclude that the distance to the Cygnus Loop is close to
the upper limit given by the distance to the background star and that 
either the electron temperatures are lower than those measured from 
{\it ROSAT}~PSPC X-ray spectral fits or an additional heat input for the
electrons, possibly due to thermal conduction, is required.
\end{abstract}

\keywords{ISM: individual (Cygnus Loop) -- shock waves -- supernova
remnants -- cosmic ray acceleration}

\section{Introduction}
Supernova explosions are responsible for showering the surrounding
interstellar medium (ISM) with heavy elements, influencing the
distribution and composition of gas in the host galaxy.  A typical
supernova releases $\sim$10$^{51}$ ergs, sending a highly supersonic
shock wave into its surroundings at speeds of up to 30,000 km~s$^{-1}$
(Ghavamian et al. 2001).  This expanding surge of energy heats and
ionizes the neighboring gas out to hundreds of parsecs, initiating
cloud collapse which leads to star formation and galactic evolution.
Although a supernova is short-lived, the ejected remains propagate
outward and interact with the ISM.  These supernova remnants (SNRs)
radiate over a large spectral range, transporting information about
the blast wave morphology and local ISM.  

The Cygnus Loop is a well-known SNR whose distance is less than
576 $\pm$ 61 pc based on the distance to a star whose spectrum shows
absorption features from shocked gas in the remnant (Blair et al., 2008).
From a global perspective, its
structure is governed by interactions with the surrounding
inhomogeneous ISM.  Despite its apparent spherical symmetry, both
optical and X-ray observations suggest that the Cygnus Loop does not
follow the simplified Sedov-Taylor morphology of an adiabatic blast
wave expanding into a homogeneous, low density medium.  Instead, the
blast wave deceleration is the result of encounters with dense,
extended clouds distributed inhomogeneously in the pre-shocked gas
(Levenson et al., 1997; Levenson et al., 1998).

We focus on the non-radiative H$\alpha$ shock filaments, primarily in
the northeastern Cygnus Loop, several of which were previously observed
by Hester et al. (1994), Blair et al. (1999, 2008), Ghavamian et
al. (2001) and Raymond et al. (2003).  In a non-radiative shock, the
cooling timescale of the hot shocked gas is much greater than the age
of the shock.  As the shocked gas has not had sufficient time to cool,
post-shock radiation does not affect the shock evolution (McKee \&
Hollenbach, 1980).  Therefore, non-radiative shock filaments provide a
laboratory for probing the conditions at the shock front before this
information is lost by radiative or collisional processes.  When a
non-radiative shock encounters neutral gas, Balmer line emission is
produced by the excitation of neutral hydrogen atoms prior to
ionization.  The resulting H$\alpha$ filaments trace the outer edge of
the Cygnus Loop non-radiative blast wave as it expands through low
density, partially neutral gas.  These H$\alpha$ filaments are useful for proper
motion measurements of shock fronts and investigations of shocked gas
parameters.

SNR shock waves are believed to accelerate cosmic rays up to energies of 
10$^{15}$ eV and to be responsible for a large
fraction of interstellar cosmic rays, implying an efficient process
for particle acceleration.  The theory of diffusive shock acceleration
describes how charged particles (dominantly protons and
electrons) passing back and forth across a shock front reach high
energies, resulting in a power-law spectrum $\propto$
E$^{-2}$ for cosmic rays above $\sim$ 10$^{9}$ GeV (Blandford \&
Eichler, 1987).  This particle acceleration process requires a
collisionless medium, as frequent collisions would return the velocity profile
to Maxwellian.  

The efficiency of energy conversion in SNR shocks into
the acceleration of cosmic rays is highly dependent on shock structure
and is a topic of active research.  Theory indicates that the fraction of
the energy dissipated in a shock going into cosmic rays is bistable, being
either very small or $\sim$ 80\% (Malkov et al. 2000).  Recent
calculations of the evolution of SNR shocks show that the efficiency
starts off low and evolves gradually to somewhat smaller values than
that predicted by steady shock models (Caprioli et al. 2008).
Synchrotron emission in the radio and X-ray bands provides direct
observation of particle acceleration in SNRs; however, the majority of
the SNR energy going into particle acceleration is claimed by protons.
Our current observational ability to detect $\gamma$-rays (and
neutrinos) is limited, so we must settle for inferences of the total cosmic ray
acceleration efficiency through indirect methods.  Warren et al. (2005)
inferred shock compression ratios greater than 4 from the distance between
the blast wave and the contact discontinuity in Tycho's SNR.  Their
results suggest that the cosmic ray to gas pressure ratio, $P_{CR}/P_G > 1$,
though projection effects might allow
a smaller compression ratio (Cassam-Chenai et al. 2008).  High cosmic
ray acceleration efficiencies have also been inferred from the low
electron temperature in 1E102-72.6 (Hughes et al. 2000) and from shock
wave precursors to Balmer line filaments (Smith et al. 1994; Hester et al.
1994).

In this paper we find upper limits to 
$P_{CR}/P_{G}$ by combining measurements of post-shock electron
temperature with the shock speed given by proper motions and distance
upper limits.  Our sample
includes 18 non-radiative H$\alpha$ shock segments dominantly in the
northeastern rim of the Cygnus Loop.  This sample, although limited by
the uncertainties in proper motion, temperature, and distance,
provides constraints on $P_{CR}/P_{G}$.  Several positions
show apparently negative values of $P_{CR}$, and we discuss the uncertainties
in temperature and distance and the possibility of additional electron
heating that could account for that non-physical result.

\section{Observations and Data Reduction}
We observe regions in the northern, eastern, and western Cygnus Loop
over optical and X-ray bands.  The Palomar Observatory Sky Survery
(POSS) obtained the optical data and the {\it ROSAT} Position
Sensitive Proportional Counter (PSPC) obtained the X-ray data.  We
reduced two epochs of optical data for proper motion measurements.
The X-ray data were obtained within two years of the more recent optical epoch.
We matched WCS coordinates of all data to ensure consistency during optical 
and X-ray analysis.

\subsection{POSS Optical Data Reduction}
The POSS observed the Cygnus Loop over two epochs spanning 39.1
years. These photographic plates have been digitized by the
Space Telescope Science Institute (STScI) and are publicly archived as
the Digitized Sky Survey.  We selected the filtered surveys whose
transmission coefficient peaked near H$\alpha$
($\lambda=6562.8~\mbox{\AA}$), corresponding to surveys POSS-I E (red)
and POSS-II F (red), to ensure detection of H$\alpha$ shock filaments
along the Cygnus Loop perimeter.  We refer to the 1953 June 15
observations as POSS-I and the 1992 July 24 observations as POSS-II,
with exposure times of 45.0 and 120 minutes, respectively.  The pixel
scales of the digitized POSS-I and POSS-II plates are 1\farcs 7 and 1\farcs 0,
respectively.

We analyzed proper motions of 18 H$\alpha$ shock filaments which
define the perimeter of the Cygnus Loop.  We used IDL and
SAOImage DS9 for subsequent image analysis.  From each POSS epoch, we
hand selected a sub-image containing the filament of interest
surrounded by reference background and foreground stars based on a common WCS (J2000)
position. POSS-I and POSS-II pixel scales were resampled to 0\farcs 1
using bilinear interpolation between grid points.  Each set of
sub-images was rotated appropriately to account for precession effects
during the time between observations.  An additional rotation was
incorporated to align the shock along the image axis for
the proper motion analysis.  Two-dimensional
cross-correlation methods determined an optimal fine-tuned pixel shift
to overlap the POSS-I and POSS-II sub-images.  In the resulting images
the background stars are stationary,
leaving the propagating shock for analysis.  All image
manipulation was made relative to the POSS-I plate in an attempt to
maintain the original integrity of the data.

\subsection{ROSAT PSPC X-Ray Data Reduction}
The Cygnus Loop X-ray data we analyzed were obtained with the PSPC
detector on-board the {\it ROSAT} observatory.  The observation
timeline spans from 1992 Nov 04 to 1994 June 03 with 3 separate
pointings targeting 3 locations.  Table 1 lists the {\it ROSAT}~PSPC
observations relevant to our X-ray post-shock analysis of the
Cygnus Loop.  All observed data sets are publicly archived through the
HEASARC.

To prepare the {\it ROSAT} PSPC data for analysis, spectral regions
were extracted from the PSPC events files requiring a minimum of 1,200
counts within each region.  Source and background spectra were
extracted and truncated to Pulse Invarient (PI) channels 0-255 using
{\tt xselect} version 2.4.  We filter PSPC spectra to accept good
events and, since we will be using $\chi^2$ statistics,
we require a minimum of 20 counts per bin using the FTOOL {\tt
grppha}.   We extracted X-ray emitting regions extending from
25\arcsec\ to 100\arcsec\ behind each H$\alpha$ filament of
interest.  As discussed below, we avoided the region within 25\arcsec\ 
where the gas is significantly out of ionization equilibrium.
Background spectra were taken outside the outer shock
front and scaled by the ratio of the areas of the extraction
regions.  PSPC response matrices were obtained from the
legacy.gsfc.nasa.gov anonymous ftp archive and ancillary response
files (ARFs) were generated using the FTOOL {\tt pcarf} as outlined by
Turner (1996).  The resulting spectra were fit with XSPEC version
11.3.2 (Arnaud, 1996).  The PSPC detector is calibrated over the 0.1-2
keV soft X-ray energy band (Prieto, 1996).  All spectra were fit in
the 0.1-1.1 keV band as more energetic photons were not typically
detected.  The X-ray errors presented in this paper are at the 90\%
confidence level obtained with the XSPEC {\tt error} command unless
otherwise indicated.

\section{Analysis and Results}
We surveyed 18 shock filaments, each of length 120\arcsec, spanning
the eastern, northern, and western Cygnus Loop outer shell which is defined by its
characteristic H$\alpha$ emission.  For each filament, we measured
proper motion by comparing two epochs of POSS red digitized plates and
post-shock gas temperature from spectral fits to {\it ROSAT} PSPC
X-ray data, assuming ionization equilibrium.  
Adopting the most recent distance upper limit to the Cygnus Loop of 576 $\pm$ 61
pc (Blair et al., 2008), we constrain the ratio of
cosmic ray to gas pressure.  We discuss our
methodology, the uncertainties in proper motion, temperature and distance,
and the results below.

\subsection{Proper Motion Measurements}
We determine the proper motion of 18 regions enclosing H$\alpha$ shock
filaments primarily along the northeastern perimeter of the Cygnus
Loop.  Figures 1-6 show POSS-II images indicating the locations and
dimensions of the extracted regions.  In all images, north is at the
top and east points to the left.  Each extracted slit, outlined in
blue with a corresponding identification number, has dimensions
25\arcsec\ $\times$ 120\arcsec\ and is positioned parallel to a
non-radiative H$\alpha$ filament.  Table 2 lists right ascension and
declination relative to POSS-I for each of the 18 regions enclosing a
non-radiative filament.  Regions are

% \bigskip
\begin{figure}[h]
\begin{center}
\includegraphics[width=3.2in,angle=0]{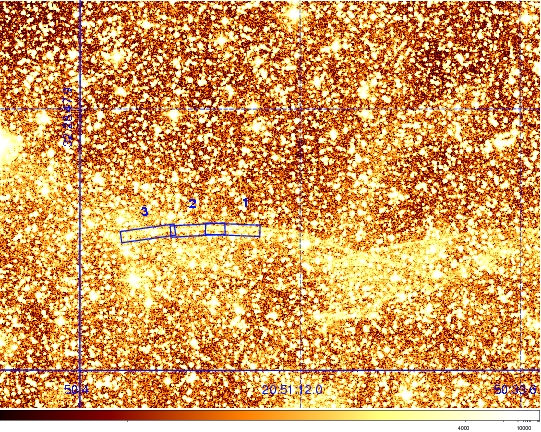}
\caption[h]{\footnotesize Image of the northern Cygnus Loop showing filaments 1-3.}
\end{center}
\label{POSS1_3.jpg}
\end{figure}
%\vspace{6 mm}

\begin{figure}[h]
\begin{center}
\includegraphics[width=3.2in,angle=0]{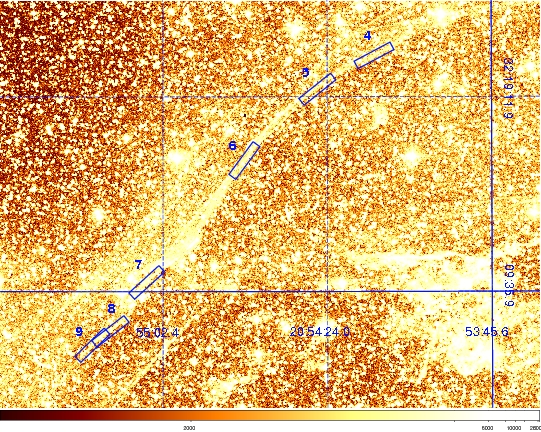}
\caption[h]{\footnotesize Image of the northeastern Cygnus Loop
showing filaments 4-9.}
\end{center}
\label{POSS4_9.jpg}
\end{figure}
%\vspace{6 mm}

\begin{figure}[h]
\begin{center}
\includegraphics[width=3.2in,angle=0]{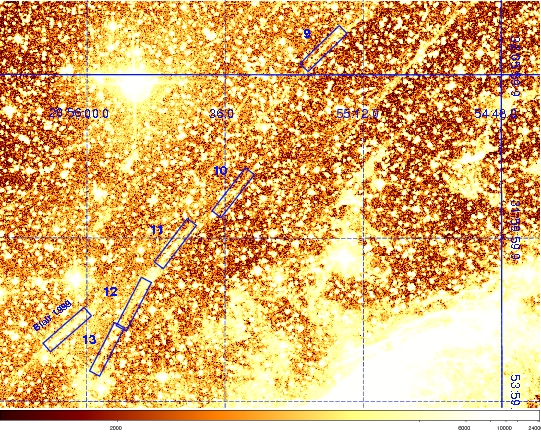}
\caption[h]{\footnotesize Image of the northeastern Cygnus Loop
showing filaments 9-13.}
\end{center}
\label{POSS10_13.jpg}
\end{figure}
%\vspace{6 mm}

\begin{figure}[h]
\begin{center}
\includegraphics[width=3.2in,angle=0]{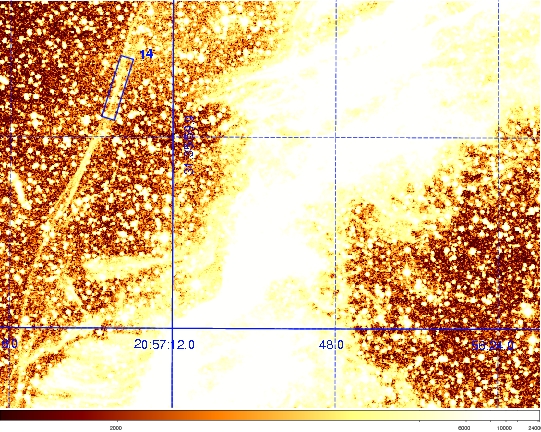}
\caption[h]{\footnotesize Image of the northeastern Cygnus Loop
showing filament 14.}
\end{center}
\label{POSS14.jpg}
\end{figure}
%\vspace{6 mm}

\begin{figure}[h]
\begin{center}
\includegraphics[width=3.2in,angle=0]{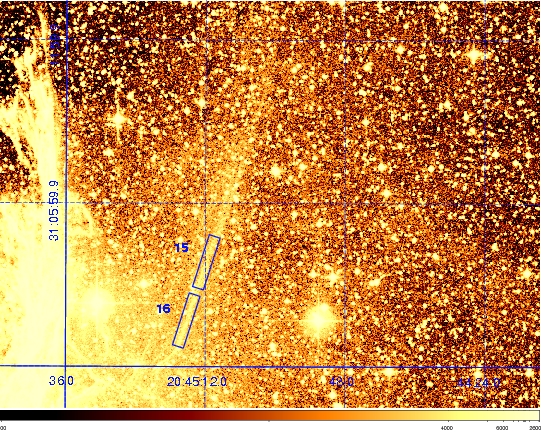}
\caption[h]{\footnotesize Image of the western Cygnus Loop
showing filaments 15 and 16.}
\end{center}
\label{POSS15_16.jpg}
\end{figure}
%\vspace{6 mm}

\begin{figure}[h]
\begin{center}
\includegraphics[width=3.2in,angle=0]{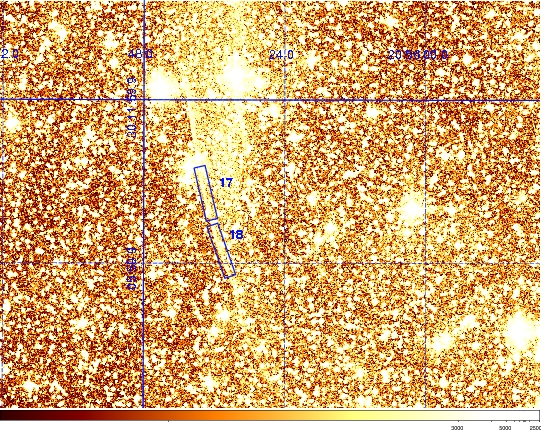}
\figcaption[h]{\footnotesize Image of the eastern Cygnus Loop
showing filaments 17 and 18.}
\end{center}
\label{POSS17_18.jpg}
\end{figure}

\noindent
labeled by a filament number which increments
with decreasing declination.  We adopted the following H$\alpha$
filament selection criteria applicable to both POSS-I and POSS-II images:

\begin{quote}
$\bullet$ Filaments must be bright relative to the local background.

$\bullet$ Selected filaments must be continuous with no local
branching structure.

$\bullet$ Filaments must not 
precede another shock wave or
strong optical emission within 100\arcsec\ to provide an extractable
{\it ROSAT}~PSPC post-shock region for X-ray temperature analysis.

$\bullet$ A region enclosing a filament is not overwhelmed by stars.

$\bullet$ Filament geometry appears to be rigid and planar over both epochs
allowing for uniform shock expansion in one dimension.
\end{quote}

\indent
Each filament is hand selected and both epochs of images are reduced
as described in $\S$ 2.1 to align the shock along the image
axis. Rectangular regions 25\arcsec\ $\times$ 120\arcsec\ parallel to
and enclosing the filament in both epochs are selected for proper
motion analysis.  Within this region we extract slices which cross the
shock but omit sub-regions containing stars.  This method of avoiding
stars ensures that the brightness of the propagating shock will
dominate its surroundings in the selected sub-regions.  Figure 7 shows
a typical example of the region selection process just described.  The
black rectangular regions of length 25\arcsec\ track shock propagation
and omit stars.  These black regions are stacked while regions
enclosed by white bars are discarded to collect all starless
sub-regions.  A bootstrap method randomly selects half of the rows
within these
starless sub-regions to sample over the length of the filament.  For
both POSS epochs, pixel values are summed down each column of sampled
slices parallel to the shock, covering the 25\arcsec\ region width.  A
one-dimensional cross-correlation of the totaled rows between both
POSS epochs determines an optimal offset corresponding to the shock
propagation.  The bootstrap process just described is repeated 1000
times to provide a proper motion distribution and a statistical error
estimate.  The number of starless pixel rows did not exceed 600 with a
required minimum of 200, ensuring well over 1000 possible combinations
for bootstrap analysis.  Figures 8 and 9 show an example of a proper
motion distribution after 1000 trials and a plot of best-fit
proper motions for a single iteration based on maximization of the
correlation coefficient, respectively.
%\bigskip

%\noindent

Table 2 lists proper motion measurements for each filament with 90$\%$
confidence errors according to the standard deviation of the randomly
sampled distribution (see Figure 8).  The time elapsed between POSS-I
and POSS-II observations is 14,284 days (39.1 years).  Proper motions
over this timeline range from 2\farcs 7 to 5\farcs 4.  We also give 
the conservative upper limit to the shock speed, $v_s$, obtained from
the upper limit to the proper motion (proper motion plus uncertainty)
and the upper limit to the distance of the background star given by
Blair et al. (2008) of 576+61 = 637 pc.  Matching
filament locations ($\alpha$, $\delta$) with corresponding proper
motions in Figures 1-6 demonstrates continuity of shock propagation
along the rim of the Cygnus Loop and consistency of our measurements.
Suitable filaments were restricted to the inner perimeter of H$\alpha$
emission to permit extraction of post-shock regions from {\it ROSAT}
X-ray data.  The post-shock gas must not
\vspace{1 mm}

\begin{figure}[h]
\begin{center}
\includegraphics[width=3.3in,angle=0]{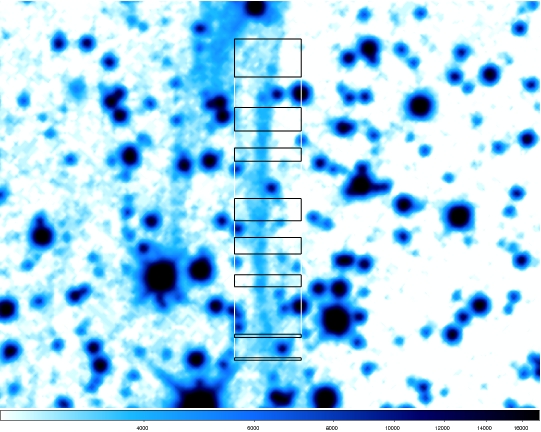}
\caption[h]{\footnotesize Typical region selection process across
an H$\alpha$ filament.  The shock is aligned along the image axis and
enclosed by a 25\arcsec\ $\times$ 120\arcsec\ rectangular region.  The black
sub-regions are selected for analysis.  The sub-regions enclosed by the white
lines are dominated by stars; therefore, they are not extracted for
proper motion analysis.  Identical sub-regions are extracted for both
POSS epochs and the columns of the remaining starless regions are
totaled for cross-correlation comparisons.  This selection process is
specific to filament 7.}
\end{center}
\label{pickregions.jpg}
\end{figure}
\vspace{5 mm}

\noindent
encounter a trailing blast wave within $\sim$ 100\arcsec\ to
validate the assumption of ionization
equilibrium required by our X-ray temperature analysis.  While there
are numerous, well-defined H$\alpha$ filaments delineating the
northeastern Cygnus loop, we only measure proper motions of those
where the post-shock gas has adequate time to reach thermal and ionization
equilibrium, allowing us to constrain cosmic ray to gas pressure.
%\vspace{1 mm}

%\noindent

\subsubsection{Proper Motion Uncertainty}
The statistical uncertainties in the proper motions are determined by
the measurement procedure to be 0\farcs 1 to 0\farcs 2 for all but one
of the filaments.  Additional systematic uncertainty could arise from
the cross registration of the images from the different epochs, though
that error should be small because of the large number of stars used in
the cross correlation.  Indeed, we have compared the positions of individual
stars in the different epochs and find them to differ by considerably less
than an arcsecond. It is also possible that changes in brightness among
different regions within an unresolved filamentary structure (see Blair
et al. 1999) could cause errors in the proper motion.  We expect such
errors to be small, and measurement of 18 positions limits the impact of
any such errors.  Thus while there is a small systematic error, it is 
unimportant because the uncertainties in temperature and distance dominate.

We can also compare with other proper motion determinations.  Shull and
Hippelein (1991) reported proper motions of 1\farcs 3 to 14\farcs 6
century$^{-1}$, with the latter value for a Balmer line filament close
to our position 14, and Hester et al. (1986) found 7\arcsec\ per
century for two Balmer line filaments in the northeast, one of which
was observed by Blair et al. (1999).
Blair et al. observed a non-radiative H$\alpha$ filament in the
Cygnus Loop ($\alpha_{J2000} = 20^{h}56^{m}2^{s}.7, ~\delta_{J2000} =
31^{\circ}56\arcmin 39\farcs 1$) propagating into relatively dense neutral gas
(see Figure 3).  The filament proper motion was measured as 3\farcs 6
$\pm$ 0\farcs 5 by comparing POSS-I and {\it Hubble Space Telescope
(HST)} images taken $\sim$ 44.3 years apart using stars as stationary
reference points.  However, the two reference stars used changed

\vspace{1 mm}

\begin{figure}[h]
\begin{center}
\includegraphics[width=2.5in,angle=90]{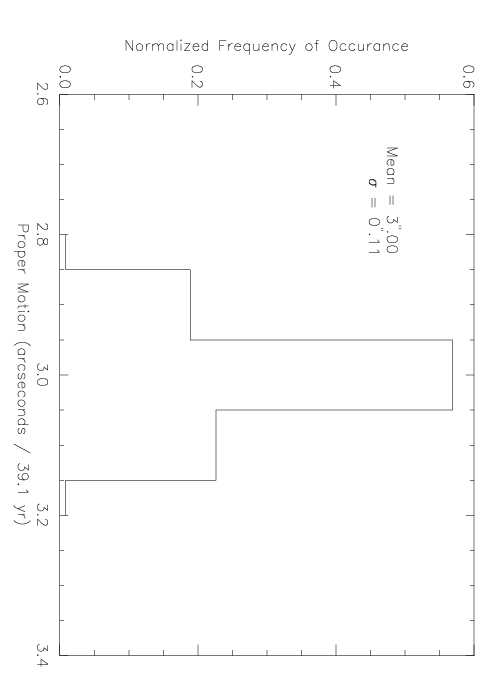}
\caption[h]{\footnotesize Histogram distribution of proper motion
measurements after 1000 bootstrap iterations.  The bins are 0\farcs 1 wide
and the results are normalized to unity.  The
90$\%$ confidence range in proper motion is represented by $\sigma$,
given by multiplying the
standard deviation by 1.645.  This distribution is specific to filament 7.}
\end{center}
\label{POSS17_18.jpg}
\end{figure}
%\vspace{5 mm}

\begin{figure}[h]
\begin{center}
\includegraphics[width=2.5in,angle=90]{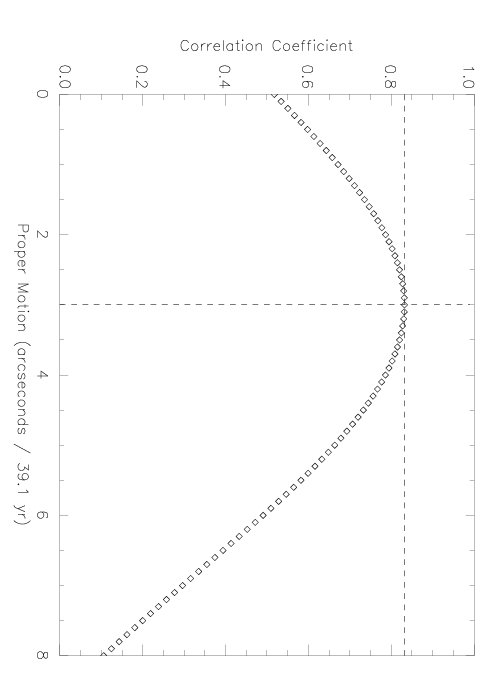}
\caption[h]{\footnotesize Spread in correlation coefficient for one
iteration between sampled POSS-I and POSS-II sub-regions contributing
to the histogram in Figure 8.  Maximization of the correlation
coefficient determines the optimal proper motion of the filament.
This distribution is specific to filament 7.}
\end{center}
\label{POSS17_18.jpg}
\end{figure}
%\vspace{5 mm}

\noindent
apparent position by 0\farcs 5 and 0\farcs 2, increasing proper motion
uncertainty.  We measured this same filament with the method described
in $\S$ 3.1 by combining POSS-I and two different POSS-II observations
taken on 1992 July 24 and 1995 August 25 yielding proper motions of
2\farcs 6 $\pm$ 0\farcs 1 and 3\farcs 1 $\pm$ 0\farcs 1, respectively.
Assuming a constant shock speed, we extrapolate these measurements to
3\farcs 0 and 3\farcs 2 at the time of the 1997 November 16 {\it HST}
observation.  These two proper motion measurements and that of Blair
et al. (1999) were made relative to the POSS-I 1953 June 15 plate,
providing three independent measurements.  Given that these results
are within the limits presented in Blair et al. (1999) when reference
star position changes are included, we consider our proper motion
methodology justified.  Although the {\it HST} image
reveals intricate substructure, if we assume that the substructure of
a particular filament has not dramatically changed over the POSS
epochs, the poor resolution of the POSS plates smoothes these features
and should give accurate results.

\subsection{Post-Shock Temperature from {\it ROSAT} Spectral Fitting}

We extracted PSPC spectra of four 25\arcsec\ $\times$
120\arcsec\ strips stacked behind and parallel to each filament for
measurements of post-shock gas temperature behavior over 100\arcsec.
The single-temperature models {\tt apec} and {\tt raymond} show low
post-shock temperature in the first $\sim$ 25\arcsec\ 
behind each filament.  Within the 100\arcsec\ post-shock region,
the fitted temperature is consistently lower within the 25\arcsec\ region
immediately behind the shock in agreement with Raymond et al. (2003).
This suggests departure from ionization equilibrium immediately
following the shock because the gas has had insufficient time to
ionize.  

We show below that the 
post-shock gas reaches ionization equilibrium after 25\arcsec, so we
extracted PSPC X-ray spectra of a 75\arcsec\ $\times$ 120\arcsec\ region
extending from 25\arcsec\ to 100\arcsec\ behind each filament.
Corresponding background regions were selected outside of the X-ray
shock front and scaled, and source and
background regions reduced as discussed above ($\S$ 2.2).  Figures 10-12
show the locations of extracted PSPC X-ray regions for all of the
H$\alpha$ filaments we considered, where north is at the top and east is
toward the left.  
% All post-shock extraction regions have dimensions 75\arcsec\ 
% $\times$ 120\arcsec.

All of our spectral fits  used the XSPEC model {\tt phabs} to account for
absorption of interstellar gas along the line of sight combined with a
hot plasma X-ray emission model.  The nominal
interstellar column density value was $N_{H} = 1.5 \times 10^{20}
\rm~cm^{-2}$, but it was allowed to vary in all model fits (Raymond et al., 2003;
Decourchelle et al., 1997).  For all the soft X-ray fits presented in
this paper, $N_{H}$ is constrained within $\sim 1-5 \times 10^{20}
\rm~cm^{-2}$. The $kT_e$ and $N_{H}$ parameters in the model are highly
(and negatively) correlated, in the sense that a very small change in
the temperature yields a large change in $N_{H}$.
The fluctuations in $N_{H}$ are likely not significant.

Ghavamian et al. (2000) measured H$\alpha$ velocity line widths of
a non-radiative shock located between filaments 5 and 6 in the
northeastern Cygnus Loop and found good electron-ion equilibration,
$T_{e}/T_{i} \sim 0.7-1.0$, in the post-shock region.  We have
assumed $T_{e} = T_{i}$ throughout our analysis.

Single-temperature models and standard cosmic abundance assumptions
are challenged by claims of bimodal temperature
regions and depleted abundances (e.g. Nemes et al., 2008).  Three fits were
made to each PSPC X-ray spectrum: (i) a single-temperature fit
assuming cosmic abundances (Anders \& Grevesse, 1989),
(ii) a single-temperature fit assuming
depleted abundances, and (iii) a double-temperature fit assuming
cosmic abundances.  Depleted abundance fits fixed C, Mg, Al, Si, Ca,
Fe, and Ni to 10$\%$ cosmic values on the basis that these elements
are locked up in grains.  Other abundances were fixed to cosmic
values.  

% It is outside the scope of this paper to rigorously
% investigate the possibility of non-standard abundances.  Similarly, we
% do not delve into detailed bimodal temperature fits for this
% preliminary stage of analysis, although we acknowledge these issues.

Table 3 lists the resulting best-fit temperatures and $\chi^{2}/\nu$
values using the XSPEC models:
\begin{quote}
(i)~~~{\tt phabs} $\times$ {\tt apec},

(ii)~~{\tt phabs} $\times$ {\tt vapec},

(iii)~{\tt phabs} $\times$ {\tt (apec~+~apec)},
\end{quote}
while Table 4 lists the same parameters using the XSPEC models:
\begin{quote}
(i)~~~{\tt phabs} $\times$ {\tt raymond},

(ii)~~{\tt phabs} $\times$ {\tt vraymond},

(iii)~{\tt phabs} $\times$ {\tt (raymond~+~raymond)}.
\end{quote}

%\textbf{[Do we need both tables 3 \& 4, or could we describe the bias in words
%in the text? Something like ``Fitting the spectra using the {\tt raymond}
%model (Raymond and Smith 1977) produces a temperature lower by about
%0.15 keV in the single-temperature fits. This is due to different
%estimates of the atomic physics and which lines are included in the
%various models.'' Also need words about the depleted and 2-Temp fits.
%Does the Kolmogoroff-Smirnov (KS) test say anything about whether the
%2nd temperature component improves the fits enough to warrant the
%extra complexity?]}

Tables 3 and 4 compare the {\tt apec} and {\tt raymond}
models to determine the sensitivity of the cosmic ray to gas pressure
ratio for slightly different X-ray temperature fits. $T_{\rm cos}$
and $T_{\rm dep}$ refer to temperature fits to models (i) and (ii),
respectively.  $T_{\rm low}$ and $T_{\rm high}$ are the bimodal
temperatures from model (iii).  Figures 13-15 show typical spectral
fits for {\tt apec} models (i), (ii), and (iii).

Temperature estimates are required for calculations of the
cosmic ray to gas pressure ratio downstream.  
The double-temperature models consistently
produce better fits based on $\chi^{2}$ statistics, and  $T_{\rm low}$
remains continuous and consistent along ajacent filaments.  However,
there is no glaring physical reason to require two downstream
temperatures.  Hester et al. (1994) discuss how reflection of a blast
wave from a dense cloud will enhance the temperature, and this may apply to 
the parts of the Cygnus Loop where the optical emission is bright, but
we have avoided those complicated areas for this analysis.  As discussed
below, temperatures also depend on the assumed set of elemental
abundances.  We estimate the post-shock
electron temperature as $T_{\rm cos}$ from the  fits
assuming cosmic abundances since they best represent the data.  Lower
limits on temperature are taken as $T_{\rm low}$ from the
double-temperature models, as this gives conservative lower limits
to the gas pressure.  Cosmic ray to gas pressure is inferred and
an upper limit is calculated based on these temperatures and measured
proper motions.  Temperatures from both {\tt apec} and {\tt raymond}
models are independently analyzed for all 18 post-shock regions as a
consistency check on temperature measurements.

\subsubsection{X-Ray Temperature Uncertainty}

The derivation of temperatures from X-ray data, especially low-resolution
X-ray data such as that provided by {\it ROSAT}, is fraught with systematic
errors. In this section, we attempt to quantify some of those errors.

Our extraction regions were chosen to provide at least 2,500 good x-ray
events in each, so that the statistical errors are minimal.

We have made two assumptions in our fitting: equilibrium ionization,
and cosmic abundances.

To investigate these, we produced spectra using the shock
code of Cox and Raymond (1985) for a shock with $v_s=350~\mathrm{km~s^{-1}}$,
which is typical of the velocities shown in Table 2, and a pre-shock
density of 0.25 $\rm cm^{-3}$, which is typical of values determined from
X-ray and UV observations (e.g., Raymond et al. 2003). We then integrated
the surface brightness to distances which correspond to 25, 50, 75,
and 100\arcsec\ behind the shock. The resulting spectra were multipled
by a photoelectric absorption model with $N_{\mathrm{H}} = 1.5\times 10^{20}~\mathrm{cm^{-2}}$,
convolved with the ROSAT response, and used with the XSPEC \texttt{fakeit} command
to generate Poisson-sampled random spectra consistent with the shock
model. We did this for both cosmic and depleted abundances.

We then fit these spectra with a cosmic abundance, single temperature
\texttt{raymond} model, assuming equilibrium ionization. We find that
for the cosmic abundance cases, the fitted temperature underestimated
the true input temperature by 3\%, which is a small effect. The
first zone, within 25\arcsec\ of the shock, had a much lower fitted
temperature. Accordingly, in our fitting of the Cygnus Loop data,
we have ignored the first 25\arcsec\ bin.

Similarly, if we model a shock with depleted abundances, we find the
spectrum is harder than with cosmic abundances. This is because much
of the 1/4~keV band flux is provided by lines of Si, Mg, and Fe, while
the 3/4~keV band flux at these shock speeds and temperatures is provided
mostly by O and Ne which are undepleted even in dusty interstellar plasmas.
% [{\bf NOTE THE CHANGE HERE.  IS IT CORRECT?}] yes... rje 9/29/08
Fitting such a spectrum with a model which presumes cosmic abundances
will overestimate the temperature to produce the harder spectrum.
However, the model allows the column density
of neutrals along the line of sight to vary, and this can compensate
and even overcompensate for the excess emission in the soft band
predicted by the cosmic abundance model, so that the temperature can
be underestimated. Overall, we find that fitting
a depleted shock spectrum to a cosmic abundance equilibrium model
underestmates the temperature by about 11\%.

The large quantity of atomic physics data which go into simulations
of the spectra of hot, collisional plasmas all have uncertainties
as well. In order to assess these effects, we have fit the spectra
with both the {\tt apec} (Desai et al. 2005, Smith et al. 2001)
and {\tt raymond} (Raymond and Smith 1977, as updated) models.
There is a systematic difference in the fitted temperatures of
single-temperature, cosmic abundance models,
in the sense that the {\tt raymond} model gives a fitted temperature about
$0.15 \pm 0.05$~keV lower than the {\tt apec} model does.  This is
probably because {\tt apec} includes only emission lines for which
reliable atomic data are available, while {\tt raymond} includes
emission lines which must be present, but for which the atomic data are
more uncertain.  This mainly affects the Mg, Si, S and Fe lines in
the 1/4 keV band.

Accordingly, we discuss \textit{lower limits}\/ to the electron temperatures.
We use the low temperatures from the two-temperature, cosmic abundance fits
as conservative lower limits to the temperature and therefore to the gas
pressure.  Even so in many cases we find that they imply upper limits to 
the cosmic ray pressure which are negative, a clearly unphysical situation.

%\textbf{We should put in words here about how the upper limits were
%%derived, and why we think that's conservative.}

\subsection{Other Temperature determinations for the Cygnus Loop}

The Cygnus Loop is one of the most studied supernova remnants in
the Galaxy, and each X-ray observatory in turn has observed it. Each
observatory has its own strengths and weaknesses, and the ensemble
of results helps to constrain the systematics of any one observation.
Many of these observations cover the area we study, with special
attention on the NE rim.

Miyata et al. (2007) analyze Suzaku observations of a field in the NE.
Their superior spectral resolution in the 0.3--2~keV band shows many
lines of elements such as C, N, O, Ne and Mg. They fit two-temperature
models to the spectra and find the higher temperature component, which
is constrained by the lines in the spectrum, has $kT$ greater than
about 0.2~keV.

Tsunemi et al. (2007) analyze data from a series of XMM-Newton pointings
across the Cygnus Loop from NE to SW. They also fit two-temperature models
to narrow concentric slices of the remnant. Their lower temperature (which
they identify as the forward shock) is always greater than $kT$~=~0.2~keV.

The \textit{Chandra}\/ observatory has also observed the NE rim. Katsuda
et al. (2008) report fits to the XSPEC \texttt{vpshock} model, which allows
for nonequilibrium ionization effects, and variable abundances. They also obtain
fitted temperatures uniformly in excess of 0.2~keV.

It therefore seems that these high temperatures are quite robust features
of the X-ray emission from the NE rim of the Cygnus Loop. Chandra's high
spatial resolution, XMM-Newton's large collecting area and Suzaku's superior
spectral resolution all lead to the same conclusion we obtain from the

\begin{figure}[h]
\begin{center}
\includegraphics[width=3.1in,angle=0]{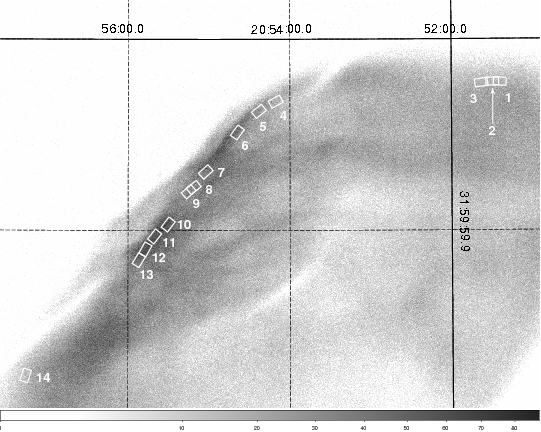}
\caption[h]{\footnotesize {\it ROSAT}~PSPC image of the northeastern
Cygnus Loop showing locations of the X-ray source
extraction regions behind filaments 1-14.}
\end{center}
\label{ROSATnorth.jpg}
\end{figure}
%\vspace{4 mm}

\begin{figure}[h]
\begin{center}
\includegraphics[width=3.1in,angle=0]{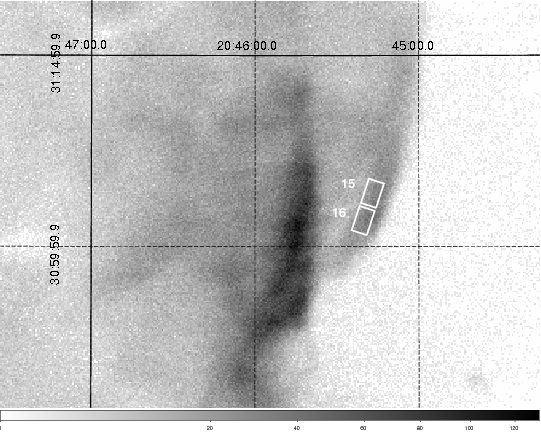}
\caption[h]{\footnotesize {\it ROSAT}~PSPC image of the western
Cygnus Loop showing locations of the X-ray source
extraction regions behind filaments 15 and 16.}
\end{center}
\label{ROSATwest.jpg}
\end{figure}
%\vspace{4 mm}

\begin{figure}[h]
\begin{center}
\includegraphics[width=3.1in,angle=0]{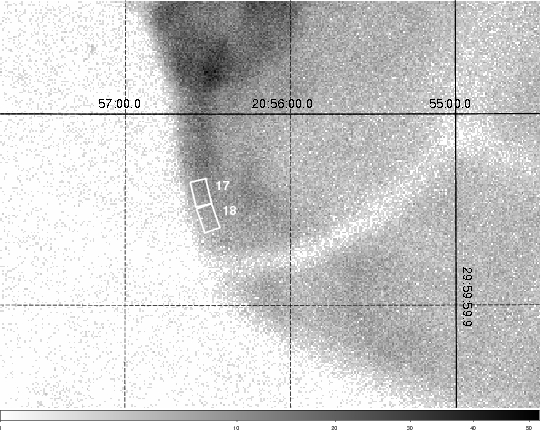}
\caption[h]{\footnotesize {\it ROSAT}~PSPC image of the eastern
Cygnus Loop showing locations of the X-ray source
extraction regions behind filaments 17 and 18.}
\end{center}
\label{ROSATeast.jpg}
\end{figure}
%\vspace{4 mm}

\begin{figure}[h]
\begin{center}
\includegraphics[width=2.2in,angle=-90]{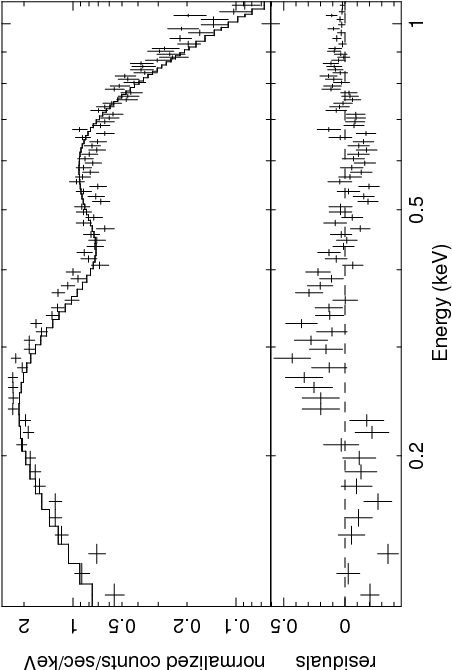}
\caption[h]{\footnotesize Representative example of a {\it
ROSAT}~PSPC single-temperature {\tt apec} model fit with abundances
fixed to cosmic values.  In most cases, this model was statistically
better than the depleted model.  This fit is specific to filament 7.}
\end{center}
\label{undepleted_apec.jpg}
\end{figure}
\vspace{5 mm}

\begin{figure}[h]
\begin{center}
\includegraphics[width=2.2in,angle=-90]{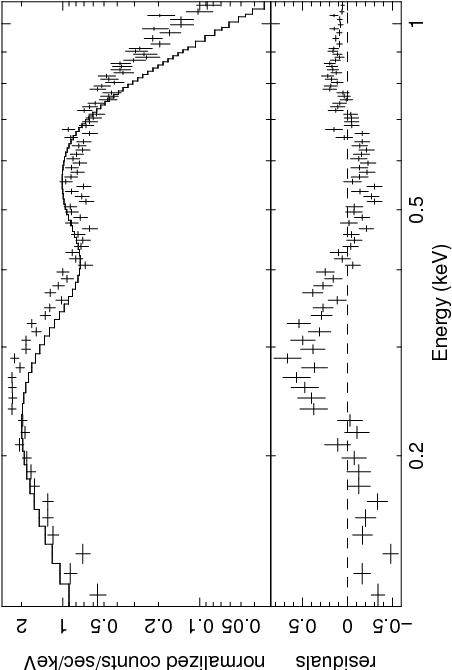}
\caption[h]{\footnotesize Representative example of a {\it
ROSAT}~PSPC single-temperature {\tt apec} model fit with C, Mg, Al,
Si, Ca, Fe, and Ni abundances depleted to 10$\%$ cosmic values and
fixed.  The rational for depletion is that these elements may be
locked up in grains.  This fit is specific to filament 7.}
\end{center}
\label{depleted_apec.jpg}
\end{figure}
%\vspace{5 mm}

\begin{figure}[h]
\begin{center}
\includegraphics[width=2.2in,angle=-90]{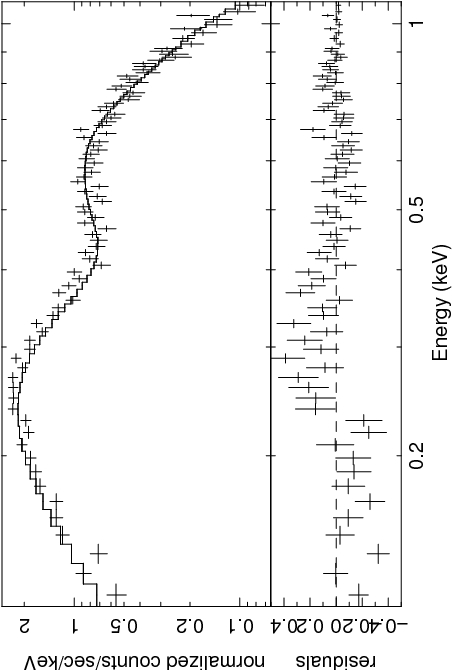}
\caption[h]{\footnotesize  Representative example of a {\it
ROSAT}~PSPC double-temperature {\tt apec} model fit with abundances
fixed to cosmic values.  In most cases, a bimodal temperature
distribution was statistically significant, providing motivation to
investigate this in future works.  This fit is specific to filament 7.}
\end{center}
\label{doubletemp_apec.jpg}
\end{figure}
%\bigskip

\noindent
{\it ROSAT} data: $kT$ values of 150~eV or higher are ubiquitous in the non-radiative
filaments of the Cygnus Loop.

\subsection{Distance}
The distance to the Cygnus Loop is a substantial uncertainty, and our
measurements can be used to obtain a lower limit to the 
distance.  Minkowski (1958) obtained the canonical value of 770 pc
from the velocity ellipse and proper motions, though Braun and Strom (1986) used the
same data to obtain 460$\pm$160 pc using the mean expansion velocity
instead of the extreme.  Shull \& Hippelein (1991) obtained
a distance of 600 pc based on Fabry-Perot scans of a large number of 
fields in the Cygnus Loop, but with a range of 300 to 1200 pc.
Sakhibov and Smirnov (1983) also compared proper motions and radial velocities
to obtain an estimate of 1400 pc.  Though the method of
comparing radial velocities and proper motions relies only upon the
assumptions of symmetric expansion and radial motion, it leads to a
wide range of distance estimates.

An independent distance estimate by Blair et al. (1999) combined the proper
motion of a single filament with a shock speed obtained from 
spectroscopic analysis and fits to shock models. They derived a distance of
440 pc with a range of 340 to 570 pc.  Their method assumed that $P_{CR}/P_{G}$
is zero, and a larger value would imply a higher shock speed and
therefore a larger distance.  This proper motion had a relatively
large uncertainty because of the possible motions of the two reference
stars.

The most solid limit on the distance to the Cygnus Loop comes from
Blair et al. (2008).  They used FUSE to observe an sdO star and found 
strong high velocity O VI absorption lines matching the O VI emission in the 
adjacent nebulosity.  A fit to the FUSE spectrum provided the temperature
and surface gravity of the star, indicating a distance of 576$\pm$61 pc.  The
upper end of this range, 637 pc, 
should be a firm upper limit to the distance of the Cygnus Loop.
Distances to sdO stars are generally difficult to determine, but the
spectral fits should give an accurate temperature.  The uncertainty
in the surface gravity probably dominates the 10\% uncertainty in the
distance.

\subsection{Analyses of other Balmer line shocks in the Cygnus Loop}

Several of the Balmer line shocks at the periphery of the Cygnus Loop have 
been studied previously.  The shock observed by Blair et al. has been
observed extensively at optical and UV wavelengths (e.g. Raymond et al.
1983; Fesen \& Itoh 1985; Long et al. 1992; Hester et al. 1994).  We do
not include it here (except to check our proper motion measurements;
$\S$ 3.1.1) because it is too slow to produce X-ray emission.  However, 
it fits in with our analysis in that the shock speed is estimated to
be 150 to 190 $\rm km~s^{-1}$ and the proper motion is around two thirds
the typical values we measure for the X-ray producing shocks.

The filament corresponding to our positions 15 and 16 was studied by
Raymond et al. (1980) and Treffers (1981), who measured an H$\alpha$
line width corresponding to a shock speed faster than 170 $\rm km~s^{-1}$.  
The filament corresponding to our positions 17 and 18 was discussed
by Fesen, Kwitter \& Downes (1992) and Graham et al. (1995).  The latter
paper finds that the Balmer line filament in the region we observed is
indeed the blast wave, rather than part of the shock structure refracted
around the dense cloud just to the north.

Detailed studies of a position in the northeast have been made
by Ghavamian et al. (2001) and Raymond et al. (2003).  We have not included
this exact region because two filaments overlap near there, but it lies
between our regions 5 and 6.  Ghavamian et al. obtained H$\alpha$ and
H$\beta$ profiles and found nearly equal proton and electron temperatures
($T_e / T_i > 0.7$ and a shock speed of 235 to 395 $\rm km~s^{-1}$.
Raymond et al. (2003) combined those results with UV spectra from FUSE
and the Hopkins Ultraviolet Telescope.  They found further evidence for
near equilibration among the particle species, with an oxygen kinetic
temperature near that of hydrogen ($T_O < 2.5 T_p$).  They also analyzed
the ROSAT X-ray spectrum, finding 0.14 $< T_e < $0.2 keV. They also 
observed a bright rim about
25\arcsec thick which they attributed to enhanced emission in the 0.25 keV
band due to non-equilibrium ionization.  They found that a 350 $\rm km~s^{-1}$
shock matches the optical, UV and X-ray data.  This speed combined with the
average of the proper motions of positions 5 and 6, 3\farcs 75, would give
a distance of 750 pc.  It should be mentioned,
however, that van Adelsberg et al. (2008) computed models of H$\alpha$
profiles in non-radiative shocks and were unable to simultaneously match
the broad line component width and the broad to narrow intensity ratio reported by
Ghavamian et al.

\section{Cosmic Ray to Gas Pressure}
The proper motion and temperature measurements described above,
combined with a distance to the Cygnus Loop, allow calculation of the
upper limit to the cosmic ray to gas pressure ratio downstream.

Taking the perspective of the shock front, momentum conservation
applies to the pre- and post-shocked gas:
\bigskip

\[\rho \frac{\partial {\bf v}}{\partial t} + \rho ({\bf v} \cdot {\bf
\nabla}) \bf{v} + {\bf \nabla}P + \rho {\bf \nabla} \phi = 0 \]

Assuming a steady solution exists, ignoring the gravitational force
term resulting from a field, $\phi$, and applying the continuity
equation $\rho v =$ constant, the momentum equation becomes,
\[\rho v \frac{d v}{d x} + \frac{d P}{d x} = \frac{d}{d x} (P + \rho
v^2) = 0. \]

This translates into a statement of constant momentum flux across the
shock, allowing us to equate pressure terms,
\[P_{1} + \frac{B_1^{2}}{8\pi} + \rho_{1} v_{s}^{2}
= P_{CR} + P_{G} + \frac{B_2^{2}}{8\pi} + \rho_{2} v_{2}^{2}, \]
where $v_{s}$ is shock speed, $\rho$ is density, $P_{CR}$ is cosmic
ray pressure originating in a dissipative shock environment, $P_{G}$
is downstream gas pressure, $B^{2}/8\pi$ is magnetic pressure, and the
subscripts 1 and 2 represent conditions in the pre- and post-shock
regions, respectively. 
% [{\bf I think we should put $B_1^2/8 \pi$ on
% the left and and change $B$ to $B_2$ on the right or else drop the
% magnetic term on the right.}] done... rje 9/29/08
We assume the ram and downstream pressures
dominate, neglecting the ambient gas pressure and interstellar
magnetic field.  Assuming the case of a strong shock with adiabatic
exponent $\gamma=\frac{5}{3}$ where $n_{2}/n_{1}=4$ and $v_{2} =
\frac{1}{4}v_{s}$, along with the equations of state $\rho_{1}=n_{1}m$
and $P_{G}=n_{2}kT_{2}$, we derive:

% $$\frac{3m v_{s}^{2}}{16kT_{2}} = 1+\frac{P_{CR}}{P_{G}}, $$

% [could we write this explicitly in terms of distance and proper motion?]
% $$\frac{3m v_{s}^{2}}{16kT_{2}} = 1+\frac{P_{CR}}{P_{G}} = \frac{3m\mu^2 d^2}{16kT_2 \Delta t^2}, $$
$$1+\frac{P_{CR}}{P_{G}} = \frac{3m v_{s}^{2}}{16kT_{2}} =  \frac{3m\mu^2 d^2}{16kT_2 \Delta t^2}, $$

\noindent
where {\it $T_{2}$} is the mean post-shock temperature.  We assume
ionization equilibrium exists, and that $T_{2} = T_{e} = T_{i}$ (Ghavamian et
al., 2000).  
We calculate shock velocities assuming only a
tangential component, $v_{s} = v_{\theta} = \frac{\mu d}{\Delta t}$,
where $\frac{\mu}{\Delta t}$ is the angular propagation across the sky
over a particular timescale (i.e. proper motion) and {\it d} is
distance to the Cygnus Loop.  
Adopting cosmic abundances with {\it n}(He)/{\it n}(H) =
0.098 (Anders \& Grevesse, 1989), the mean mass per particle behind
the shock, {\it m}, is determined by dividing the weighted molar mass
per nucleus by the average number of particles per nucleus.  The cosmic ray
to gas pressure ratios presented in this paper are computed according
to the equation derived above.  We note that the compression factor of
4 may be an underestimate if cosmic ray pressure significantly impacts
the gas dynamics, decreasing the post-shock flow speed relative to the
shock, but as will be seen, we are primarily interested in low efficiency
shocks.  
We neglect line of sight velocity components in
our calculations, assuming the shock velocity is mostly perpendicular to the
line of sight.  We note that the appearance of
multiple propagating shock layers is an artifact of our
two-dimensional perspective.  Instead, the complicated
three-dimensional structure of the Cygnus Loop is wavy, warped, and
sheet-like (Raymond et al. 2003).  Cosmic ray to gas pressure ratios
were calculated by combining proper motion and temperature
measurements with an adopted distance of 576 $\pm$ 61 pc, obtained
from a distance upper limit to a background subdwarf OB star (Blair et
al., 2008).

Note that the density has dropped out of the equations. The results
we present here do not depend upon the ambient density, which is
not very well known.

Table 2 lists best-fit values and upper limits of $P_{CR}/P_{G}$ for
all 18 filaments calculated independently with the {\tt apec} and {\tt
raymond} ionization equilibrium temperature fits.  To obtain conservative
upper limits, we have used the upper limits to the shock speed $v_s$
from Table 2 based on upper limits to the proper motions and the upper
limit to the distance of 637 pc (Blair et al. 2008).  We 
find several negative values for $P_{CR}/P_{G}$.  This result is unphysical
and requires further investigation.  For all instances of negative
pressure ratios we set $P_{CR}/P_{G}$ to zero and derive a minimum
distance estimate to the Cygnus Loop based on our proper motion and
temperature uncertainties.

\section{Discussion}
Inspection of the derived cosmic ray to gas pressure equation
emphasizes the importance of constraining proper motion, distance, and
temperature to infer tight upper limits on $P_{CR}/P_{G}$.
$P_{CR}/P_{G}$ scales with the square of the shock speed, given by the
product of proper motion and distance.  In general, the upper limits
indicate a small value for the ratio, and in some cases it is formally
negative. 

In the context of the discussions of the uncertainties given above,
it is clear that the proper motion measurements cannot account for the
unphysical values of $P_{CR}/P_{G}$.  Most distance estimates are significantly
smaller than the 637 pc we have adopted as the nominal upper limit,
and it seems like a strain to make the distance large enough ($\sim 1$ kpc)
to make all the upper limits to $P_{CR}/P_{G}$ positive.  However, it
may be possible given the general uncertainties in SdO star distances.

Therefore, we conclude that the heart of uncertainty
likely lies in post-shock temperature measurements and electron-ion
thermal equilibration assumptions.  Assuming ionization equilibrium
and that shock velocity measurements are reasonably accurate, we
require lower post-shock electron temperatures than measured from our
{\it ROSAT}~PSPC X-ray spectral fits.  The fits based on the Raymond
\& Smith (1977) code give temperatures about 10\% lower than those
using APEC, but that is several times too small to explain the
discepancy.  

\subsection{Heating by Thermal Conduction}
There is another possible way out of the discrepancy.  We have
assumed that $T_e = T_i$ behind the shock, and it is possible that
$T_e > T_i$.  However, such behavior is not seen in shocks in the
interplanetary medium, and there is no obvious physical reason for
such electron heating in the shock.  Ghavamian et al. (2006) describe
a wave heating mechanism that would produce particle temperatures
somewhat higher than observed here, but $T_p$ from Ghavamian
et al. (2001) and a higher $T_e$ would then require a large value of $V_s$.  

An interesting alternative is electron heating by thermal conduction from
gas farther behind the shock.  A number of papers have explored the 
global structure of SNRs with thermal conduction (e.g., Slavin \& Cox 
1992; Cox et al. 1999; Shelton et al. 1999).  We can obtain an
observational estimate from X-ray observations.  Nemes et al. (2008) show 
the variation of $T_e$ with radius in the northeastern Cygnus Loop based
on XMM spectra, with a gradient of 0.05 keV over a distance of 0.07 times the
shock radius, or  $\nabla T \sim 1.8 \times 10^{-13}~\rm K~cm^{-1}$.
This implies a volumetric heating rate of about $3.4 \times 10^{-22}~
\rm erg~cm^{-3}~s^{-1}$, which could heat the gas by about $3 \times 10^5$ K
as it travels over the 100\arcsec\ region we analyze.  Thus the true
post-shock temperature could be somewhat lower than the X-ray temperatures
we measure by approximately the amount required for $P_{CR}$ greater
than zero.  The above estimate requires a temperature gradient parallel
to the magnetic field, so it could not be correct everywhere.  However,
only 7 of the 18 positions show $P_{CR}$ nominally smaller than zero.

\section{Summary}
Constraining $P_{CR}/P_{G}$ in supernova
shocks is important for assessing the efficiency of energy dissipated
by the SNR into accelerating cosmic rays.  We combine measured
proper motions with temperatures derived from X-ray spectra and an
upper limit to the distance to obtain upper limits to $P_{CR}/P_{G}$.
We measured proper motions and post-shock temperatures of 18 faint
nonradiative H$\alpha$ filaments in the Cygnus Loop.  Proper motion
measurements of filaments based on image matching and correlation
techniques yield continuous results over extended shock regions.
Post-shock electron temperatures from X-ray fits are higher than
expected if the current 576 $\pm$ 61 pc distance measurement to the
Cygnus Loop is accurate.  

We conclude that 1) $P_{CR}/P_{G}$ is small, 2) the distance to the
Cygnus Loop must be close to or even larger than the upper limit given
by the apparent distance to a star that lies behind the SNR,
3) uncertainties in the temperature derived from X-ray spectra might
dominate the uncertainties in the analysis, and 4) thermal conduction
from hotter interior gas could alter the immediate post-shock temperature
enough to account for the observations.

Future work will focus on constraining
post-shock temperature and investigating the validity of ionization
equilibrium on a more global scale along the northeastern Cygnus Loop,
and placing tighter constraints on the distance to the Cygnus Loop.
\vspace{8 mm}

\noindent
% GS is happy to extend thanks to John Raymond and Dick Edgar for their
% highly enthusiastic guidance and most importantly their patience all
% summer.  These are rare qualities which all advisors should strive
% toward.  Being involved in exciting research with John and Dick
% solidified and motivated my hopes for a career in astronomy far more
% than any academic courses have.  I also wish to thank Marie Machacek
% and Jonathan McDowell for their genuine friendliness and caring
% attitudes which made my research experience very pleasant.  Finally,
% thanks are certainly due to all individuals involved in the SAO Summer
% Intern Program made possible by a generous NSF grant. 
This work is supported in part by the National Science Foundation
Research Experiences for Undergraduates (REU) and Department of Defense
Awards to Stimulate and Support Undergraduate Research Experiences (ASSURE)
programs under Grant no. 0754568 and by the Smithsonian Institution.

We acknowledge generous data policies of the Space Telescope Science
Institute, for digitizing and archiving the Palomar Observatory Sky
Surveys, and the High Energy Astrophysics Space Astronomy Archival
Research Center (HEASARC) at the NASA/Goddard Space Flight Center,
for making the ROSAT data available.
The Digitized Sky Survey was produced at the Space Telescope Science Institute under U.S. Government grant NAG W-2166. The images of these surveys are based on photographic data obtained using the Oschin Schmidt Telescope on Palomar Mountain and the UK Schmidt Telescope. The plates were processed into the present compressed digital form with the permission of these institutions.
RJE acknowledges
support from NASA contract NAS8-03060 (the {\it Chandra} X-ray Center)
to the Smithsonian Astrophysical Observatory.

\vspace{100 mm}

\begin{figure}[h]
\begin{center}
\includegraphics[width=4.5in,angle=90]{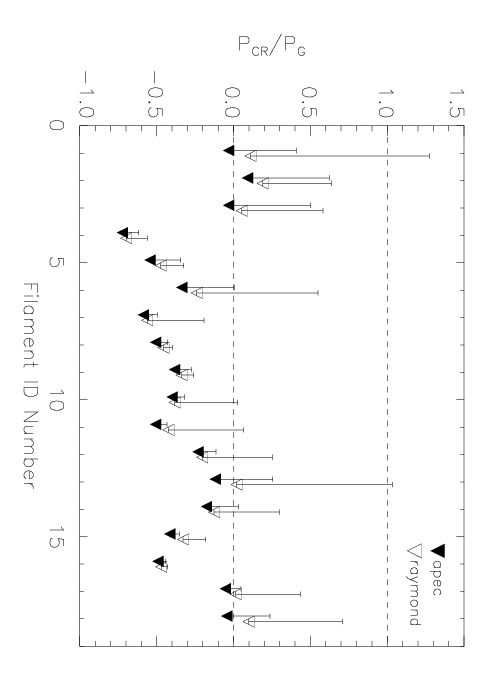}
%\centerline{~\psfig{file=my_PcrPg.ps,width=5.0in,angle=270}~}
\caption[h]{\footnotesize Mapping filaments by their numbers
defined in Figures 1-6 corresponding to $P_{CR}/P_{G}$ best fits with
error bars extending to the calculated upper limit.  Results for XSPEC
model fits {\tt apec} are shown as filled symbols and {\tt raymond} results are
shown as open symbols.  Groupings of local filament regions cluster around similar
$P_{CR}/P_{G}$ values.  Filaments 4-13, which delineate the
northeastern rim of the Cygnus Loop, appear to show $P_{CR}/P_{G}$
values consistent with zero.}
\end{center}
\label{PcrPg_recalc.jpg}
\end{figure}
%\vspace{5 mm}

%\centerline{~\psfig{file=PcrPg_ratio_recalc.ps,width=5.7in,angle=90}~}
%\figcaption[h]{\footnotesize Comparison of $P_{CR}/P_{G}$ best-fits (*) and
%upper limits ($\triangle$) calculated from {\tt apec} and {\tt raymond} XSPEC
%diffuse gas models.  Linear fits to the $P_{CR}/P_{G}$ best fits and
%upper limits
%are based on $\chi^{2}$ minimization.  Slopes of the fits are
%represented by {\it m}.  Slopes approaching unity imply good agreement
%between {\tt apec} and {\tt raymond} models.  Upper limits on
%$P_{CR}/P_{G}$ seem to be strongly model dependent which is not the
%case for best-fit calculations.}
%\bigskip

%\centerline{~\psfig{file=cyg_loop_obs.ps,width=7.5in,angle=0}~}
%\figcaption[h]{\footnotesize Collection of archived X-ray telescope
%pointings overlayed atop the Cygnus Loop.  Region sizes correspond
%roughly to detector field area.  ${\rm Red}={\it ROSAT}$, ${\rm Yellow}={\it
%Suzaku}$, ${\rm Green}={\it Chandra}$, and ${\rm Blue}={\it XMM-Newton}$.
%{\bf Do we need this figure?}}
%
%
%
%
\begin{table}[ht]
\caption~\center{Cygnus Loop {\it ROSAT} ~PSPC Observation Parameters}
\begin{center}
\begin{footnotesize}
\begin{tabular}{c c c c c}
\hline
\hline \\

Name  &  $\alpha_{J2000}$  &  $\delta_{J2000}$  &  Date  &  Exposure  \\
      &   &   &  (UT)  &  (ksec)    \\
\hline
\\

rp500034a01  &  20 55 02.4  &  +32 07 12.0  &  1992 Nov 04  &  12.0  \\
\\
rp500267n00  &  20 56 21.6  &  +30 24 00.0  &  1993 Nov 14  &  9.24  \\
\\
rp500268n00  &  20 45 40.8  &  +31 02 24.0  &  1994 Jun 03  &  9.75  \\

\\
\hline
\end{tabular}

\tablecomments{The coordinates listed describe the target location for a particular {\it
ROSAT}~PSPC pointing.  Units of right ascension are hours, minutes, and
seconds.  Units of declination are degrees, arcminutes, and
arcseconds.  We adopt this system of units throughout this paper.}

\vspace{-1.0\baselineskip}
\end{footnotesize}
\end{center}
\end{table}

\begin{table}[ht]
\caption~\center{Proper Motion and $P_{cr}/P_{g}$ Values for Selected
H$\alpha$ Filaments}
\begin{center}
\begin{footnotesize}
\begin{tabular}{c c c c c c c c c c c}
\hline
\hline \\

Filament ID  &  $\alpha_{J2000}$  &  $\delta_{J2000}$  &  Proper
Motion  &  $v_{max}$  &  $(P_{CR}/P_{G})_{\rm apec}$  &
$(P_{CR}/P_{G})^{MAX}_{\rm apec}$  & d$^{\rm MIN}_{\rm apec}$  &
$(P_{CR}/P_{G})_{\rm raymond}$  & $(P_{CR}/P_{G})^{MAX}_{\rm raymond}$
& d$^{\rm MIN}_{\rm raymond}$  \\

% Units
&   &   &  arcsec / 39.1 yr   &  km~s$^{-1}$  &  &  &  pc  &  &
&  pc  \\

\hline
\\

%15
1  & 20 51 23.9 & 32 24 22.5 &  5\farcs 1~$\pm$~0\farcs 2  &  403 
  &  -0.032  &  0.408  &  537  &  0.111  &  1.274  &  422   \\
\\
%16
2  & 20 51 29.9 & 32 24 21.8 &  5\farcs 4~$\pm$~0\farcs 2  &  433
  &  0.091  &  0.625  &  500  &  0.190  &  0.637  &  498   \\
\\
%18
3  & 20 51 38.6 & 32 24 14.5 &  5\farcs 2~$\pm$~0\farcs 2  &  416
  &  -0.032  &  0.501  &  520  &  0.053  &  0.582  &  507   \\
\\
%14
4  & 20 54 13.1 & 32 21 14.0 &  2\farcs 7~$\pm$~0\farcs 2  &  225  
  &  -0.723  &  -0.616  &  1028  &  -0.698  &  -0.560  &  960  \\
\\
%11
5  & 20 54 26.2 & 32 19 36.5 &  3\farcs 4~$\pm$~0\farcs 2  &  278  
  &  -0.543  &  -0.344  &  786  &  -0.473  &  -0.325  &  775  \\
\\
%10
6  & 20 54 43.4 & 32 16 04.0 &  4\farcs 1~$\pm$~0\farcs 2  &  333  
  &  -0.337  &  -0.004  &  638  &  -0.239  &  0.550  &  512  \\
\\
%7
7  & 20 55 06.3 & 32 10 03.8 &  3\farcs 0~$\pm$~0\farcs 1  &  240  
  &  -0.587  &  -0.496  &  897  &  -0.561  &  -0.193  &  709  \\
\\
%6
8  & 20 55 14.7 & 32 07 38.9 &  3\farcs 4~$\pm$~0\farcs 1  &  274  
  &  -0.507  &  -0.428  &  843  &  -0.456  &  -0.397  &  821  \\
\\
%5
9  & 20 55 18.9 & 32 06 58.0 &  3\farcs 7~$\pm$~0\farcs 1  &  294  
  &  -0.384  &  -0.275  &  748  &  -0.338  &  -0.259  &  740  \\
\\
%4
10 & 20 55 34.6 & 32 01 40.2 &  3\farcs 5~$\pm$~0\farcs 1  &  279  
  &  -0.400  &  -0.317  &  771  &  -0.383  &   0.026  &  629  \\
\\
%3
11 & 20 55 44.9 & 31 59 43.8 &  3\farcs 1~$\pm$~0\farcs 2  &  254  
   &  -0.507  &  -0.432  &  845  &  -0.424  &  0.066  &  617    \\  
\\
%2
12 & 20 55 51.8 & 31 57 34.8 &  4\farcs 0~$\pm$~0\farcs 2  &  319  
  &  -0.231  &  -0.115  &  677  &  -0.203  &  0.253  &  569  \\
\\
%1
13 & 20 55 56.5 & 31 55 55.1 &  4\farcs 2~$\pm$~0\farcs 7  &  375  
  &  -0.120  &  0.256  &  568  &  0.020  &  1.034  &  447  \\  
\\
%20
14 & 20 57 20.2 & 31 37 32.4 &  4\farcs 3~$\pm$~0\farcs 1  &  342  
  &  -0.176  &  0.030  &  628  &  -0.127  &  0.299  &  559  \\
\\
%28
15 & 20 45 11.9 & 31 03 50.0 &  3\farcs 4~$\pm$~0\farcs 1  &  272  
  &  -0.415  &  -0.351  &  791  &  -0.329  &  -0.182  &  704  \\
\\
%29
16 & 20 45 15.3 & 31 01 41.4 &  3\farcs 3~$\pm$~0\farcs 1  &  264  
  &  -0.492  &  -0.443  &  854  &  -0.470  &  -0.428  &  842  \\
\\
%24
17 & 20 56 37.4 & 30 08 34.4 &  4\farcs 5~$\pm$~0\farcs 1  &  358  
 &  -0.054  &  0.049  &  622  &  0.013  &  0.434  &  532  \\
\\
%25
18 & 20 56 34.8 & 30 06 27.8 &  4\farcs 8~$\pm$~0\farcs 1  &  379  
 &  -0.044  &  0.237  &  573  &  0.099  &  0.707  &  ...  \\

\\
\hline
\end{tabular}

\tablecomments{The coordinates listed represent the right ascension
and declination at the center of the extracted filament.  The measured
proper motions are derived from comparing POSS-I and POSS-II images
observed 39.1 years apart.  Errors on proper motion do not include a
$\le$ 0\farcs 1 uncertainty from image alignment.  Shock speed, $v_{s}$, is
calculated from the product of proper motion and distance using the upper
limits of proper motion + uncertainty and 576+61 = 637 pc (Blair et al. 2008).
All cosmic ray to gas pressure ratio calculations are based on these
conservative upper limits.
Minimum distances to the Cygnus Loop based
on our measurements assume $P_{CR}/P_{G} = 0$ with an upper limit on
proper motion and lower limit on temperature.  We compare results from
individual X-ray temperature fits with the XSPEC models {\tt apec} and
{\tt raymond}.
%[{\bf what to do with raymond max P/P ratio, min distance for 18}]
}

\vspace{-1.0\baselineskip}
\end{footnotesize}
\end{center}
\end{table}

%--------------------------------------------------------------

\begin{table}[ht]
\caption~\center{X-ray Spectral Fit Parameters with the {\tt apec} Model}
\begin{center}
\begin{footnotesize}
\begin{tabular}{c c c c c c c c c}
\hline
\hline \\

%APEC

Filament ID  &  T$_{\rm cos}$  &  N$_{\rm H,cos}$ &  $(\chi^2/\nu)_{\rm cos}$  &  T$_{\rm dep}$  &
$(\chi^2/\nu)_{\rm dep}$ &  T$_{\rm low}$  &  T$_{\rm high}$   &
$(\chi^2/\nu)_{\rm double}$  \\

% Units
  &  eV  &   $10^{20} \rm~cm^{-2}$ & &  eV  &  &  eV  &  eV  &  \\

\hline
\\

%15
1  &  187$^{+6}_{-6}$  &  2.4$^{+0.2}_{-0.2}$  &   1.65  &  162$^{+10}_{-7}$  &  3.37
   &  137  &  536  &  1.13  \\
\\
%16
2  &  192$^{+7}_{-6}$  &  2.4$^{+0.2}_{-0.2}$  &  1.80  &  198$^{+10}_{-9}$  &  3.01
   &  137  &  602  &  1.28  \\
\\
%18
3  &  198$^{+7}_{-6}$  &  2.5$^{+0.2}_{-0.2}$  &  1.44  &  230$^{+10}_{-10}$  &  2.47
   &  137  &  336  &  1.16  \\
\\
%14
4  &  190$^{+6}_{-6}$  &  2.3$^{+0.2}_{-0.2}$  &  0.96  &  194$^{+9}_{-8}$  &  1.86
   &  157  &  454  &  0.85  \\
\\
%11
5  &  181$^{+5}_{-4}$  &  2.5$^{+0.2}_{-0.2}$  &   1.48  &  159$^{+8}_{-6}$  &  3.21
   &  140  &  605  &  1.06  \\
\\
%10
6  &  179$^{+4}_{-2}$  &  2.1$^{+0.1}_{-0.1}$  &  2.33  &  151  &  5.75
   &  132  &  380  &  1.60  \\
\\
%7
7  &  154$^{+5}_{-4}$  &  2.6$^{+0.2}_{-0.2}$  & 2.22  &  119  &  5.95
   &  136  &  541  &  1.53  \\
\\
%6
8  &  170$^{+12}_{-4}$  &  1.2  &  1.26  &  118$^{+4}_{-4}$  &  3.44
   &  156  &  661  &  1.06  \\
\\
%5
9  &  159$^{+6}_{-5}$  &  1.7$^{+0.2}_{-0.2}$  &  1.45  &  115$^{+4}_{-4}$  &  3.63
   &  141  &  656  &  1.14  \\
\\
%4
10 &  146  &  2.1  &  5.59  &  80.8  &  10.7
   &  135  &  ...  &  3.90  \\
\\
%3
11 &  139  &  1.9  &  5.07  &  80.8  &  9.92
   &  135  &  690  &  3.95  \\
\\
%2
12 &  145$^{+3}_{-3}$  & 1.1$^{+0.1}_{-0.1}$  &  2.91  &  80.8  &  7.18
   &  136  &  685  &  1.55  \\
\\
%1
13 &  139  &  1.7  &  5.03  &  80.8  &  9.34
   &  133  &  863  &  3.40  \\
\\
%20
14 &  159$^{+8}_{-7}$  &  1.8$^{+0.3}_{-0.3}$  &   1.48  &  115$^{+7}_{-4}$  &  3.14
   &  135  &  727  &  0.89  \\
\\
%28
15 &  140$^{+3}_{-5}$  &  2.1$^{+0.2}_{-0.2}$  &  2.81  &  80.8  &  5.24
   &  135  &  812  &  2.29  \\
\\
%29
16 &  150$^{+4}_{-4}$  &  2.4$^{+0.1}_{-0.1}$  &  1.02  &  86.8  &  1.91
   &  148  &  692  &  1.03  \\
\\
%24
17 &  152$^{+13}_{-9}$  &  4.2$^{+0.6}_{-0.7}$  &  1.19  &  146$^{+9}_{-7}$  &  1.91
   &  145  &  159  &  1.24  \\
\\
%25
18 &  169$^{+11}_{-20}$  &  3.8$^{1.1}_{-0.5}$  &   1.32  &  155$^{+23}_{-10}$  &  1.88
   &  138  &  633  &  1.24  \\

\\
\hline
\end{tabular}

\tablecomments{{\it ROSAT}~PSPC X-ray spectral parameters corresponding to fits behind
each filament with the XSPEC model {\tt apec}.  The subscript 'cos'
refers to single-temperature model fits {\tt phabs} $\times$ {\tt
apec} with  abundances fixed to cosmic.  The subscript 'dep'
refers to single-temperature model fits {\tt phabs} $\times$ {\tt
vapec} with depleted abundances fixed to 10$\%$ cosmic.  The subscript 'double'
refers to double-temperature model fits {\tt phabs} $\times$ {\tt
(apec+apec)} with  abundances fixed to cosmic.  We allowed
$N_{H}$ to vary from an initial value of $ 1.5 \times 10^{20} \rm~cm^{-2}$.  We
stress that the errors listed in the table are generated in XSPEC
and are not representative of actual uncertainties based on the
variation in best fit values when comparing similar models.  Fits with
no errors or parameters listed are unphysical or limited by
$\chi^{2}$ statistics.}

\vspace{-1.0\baselineskip}
\end{footnotesize}
\end{center}
\end{table}

%---------------------------------------------------------------------

\begin{table}[ht]
\caption~\center{X-ray Spectral Fit Parameters with the {\tt raymond} Model}
\begin{center}
\begin{footnotesize}
\begin{tabular}{c c c c c c c c c}
\hline
\hline \\

%RAYMOND

Filament ID  &  T$_{\rm cos}$  &  N$_{\rm H,cos}$ &  $(\chi^2/\nu)_{\rm cos}$  &  T$_{\rm dep}$  &
$(\chi^2/\nu)_{\rm dep}$ &  T$_{\rm low}$  &  T$_{\rm high}$   &
$(\chi^2/\nu)_{\rm double}$ \\

% Units
  &  eV  &  $10^{20} \rm~cm^{-2}$  &  &  eV  &  &  eV  &  eV  &  \\

\hline
\\

%15
1  &  163$^{+7}_{-8}$  & 3.5$^{+0.5}_{-0.4}$  &   1.69  &  453$^{+22}_{-20}$  &  1.57
   &  84.8  &  360  &  0.93  \\
\\
%16
2  &  176$^{+8}_{-3}$  & 3.0$^{+0.2}_{-0.4}$  &  1.89  &  485$^{+27}_{-24}$  &  1.70
   &  136  &  702  &  1.27  \\
\\
%18
3  &  182$^{+6}_{-5}$  & 2.9$^{+0.2}_{-0.2}$  &  1.47  &  487$^{+26}_{-24}$  &  2.31
   &  130  &  356  &  1.13  \\
\\
%14
4  &  174$^{+6}_{-11}$  & 2.8$^{+0.6}_{-0.3}$  &   1.03  &  444$^{+20}_{-16}$  &  1.93
   &  137  &  448  &  0.84  \\
\\
%11
5  &  157$^{+6}_{-5}$  &  3.6$^{+0.4}_{-0.3}$  &  1.51  &  432$^{+10}_{-10}$  &  2.19
   &  136  &  776  &  1.06  \\
\\
%10
6  &  156$^{+4}_{-4}$  &  3.2$^{+0.2}_{-0.2}$  &  2.17  &  427$^{+9}_{-15}$  &  3.18
   &  84.8  &  266  &  1.28   \\
\\
%7
7  &  145$^{+3}_{-2}$  &  3.5$^{+0.2}_{-0.2}$  &  2.02  &  372$^{+12}_{-12}$  &  2.85
   &  84.9  &  292  &  1.09  \\
\\
%6
8  &  154$^{+4}_{-4}$  &  2.1$^{+0.2}_{-0.2}$  &  1.16  &  430$^{+6}_{-9}$  &  2.29
   &  148  &  ...  &  1.08  \\
\\
%5
9  &  148$^{+4}_{-4}$  &  2.5$^{+0.2}_{-0.2}$  &  1.34  &  415$^{+17}_{-17}$  &  2.02
   &  138  &  ...  &  1.11  \\
\\
%4
10 &  142  &  2.8  &  5.19  &  373$^{+10}_{-9}$  &  3.23
   &  89.9  &  539  &  2.60  \\
\\
%3
11 &  119$^{+5}_{-5}$  &  3.6$^{+0.4}_{-0.3}$  &  4.61  &  326$^{+8}_{-8}$  &  2.32
   &  71.9  &  449  &  1.67  \\
\\
%2
12 &  140$^{+3}_{-1}$  &  1.9$^{+0.1}_{-0.2}$  &  2.58  &  379$^{+10}_{-9}$  &  2.08
   &  96.1  &  425  &  1.04  \\
\\
%1
13 &  120$^{+5}_{-5}$  &  3.2$^{+0.4}_{-0.3}$  &  4.63  &  366$^{+11}_{-10}$  &  2.98
   &  82.1  &  627  &  1.84  \\
\\
%20
14 &  150$^{+6}_{-5}$  &  2.6$^{+0.3}_{-0.3}$  &  1.34  &  431$^{+21}_{-23}$  &  1.17
   &  107  &  670  &  0.84  \\
\\
%28
15 &  122$^{+9}_{-7}$  &  3.7$^{+0.6}_{-0.5}$  &  2.66  &  114$^{+4}_{-5}$  &  2.83
   &  107  &  704  &  1.81  \\
\\
%29
16 &  144$^{+4}_{-3}$  &  0.9$^{+0.2}_{-0.1}$  &  1.02  &  133$^{+3}_{-5}$  &  1.16
   &  144  &  145  &  1.06  \\
\\
%24
17 &  142$^{+6}_{-5}$  &  5.3$^{+0.5}_{-0.6}$  &  1.22  &  143$^{+9}_{-6}$  &  1.39
   &  106  &  693  &  1.03  \\
\\
%25
18 &  147$^{+10}_{-7}$  &  5.2$^{0.7}_{-0.8}$  &  1.32  &  150$^{+12}_{-10}$  &  1.52
   &  ...  &  147  &  1.39  \\

\\
\hline
\end{tabular}

\tablecomments{{\it ROSAT}~PSPC X-ray spectral parameters corresponding to fits behind
each filament with the XSPEC model {\tt raymond}.  The subscript 'cos'
refers to single-temperature model fits {\tt phabs} $\times$ {\tt
raymond} with  abundances fixed to cosmic.  The subscript 'dep'
refers to single-temperature model fits {\tt phabs} $\times$ {\tt
vraymond} with depleted abundances fixed to 10$\%$ cosmic.  The subscript 'double'
refers to double-temperature model fits {\tt phabs} $\times$ {\tt
(raymond+raymond)} with  abundances fixed to cosmic.  We allowed
$N_{\rm H}$ to vary from an initial value of  $ 1.5 \times 10^{20} \rm~cm^{-2}$.  We
stress that the errors listed in the table are generated in XSPEC
and are not representative of actual uncertainties based on the
variation in best fit values when comparing similar models.
Fits with no errors or parameters listed are unphysical or limited by
$\chi^{2}$ statistics.}

\vspace{-1.0\baselineskip}
\end{footnotesize}
\end{center}
\end{table}


\begin{references}

\reference{} Anders, E., Grevesse, N., 1989, GeCoA, 53, 197A

\reference{} Arnaud, K. A., 1996, ADASS V, eds. G. Jacoby, \&
J. Barnes, ASP Conf.\ Series 101, 17

\reference{} Beuermann, K., 2008, A\&A, 481, 919

\reference{} Blair, W., Sankrit, R., Raymond, J. C., \& Long, K. S.,
1999, AJ, 118, 942

\reference{} Blair, W. P., Sankrit, R., Torres, S. I., Chayer, P.
\& Danforth, C. W., 2008, ApJ, submitted.

\reference{} Blandford, R., \& Eichler, D., 1987, Physics Reports,
154, 1

\reference{} Boulares, A., \& Cox, D. P., 1988, ApJ, 333, 198

\reference{} Braun, R. \& Strom, R.G. 1986, A\&A, 164, 208

\reference{} Cox, D.P., \& Raymond, J. C. 1985, ApJ, 298, 651

\reference{} Cox, D.P., Shelton, R.L., Maciejewski, W., Smith, R.K.,
Plewa, T., Pawl, A. \& R\'{o}\.{z}yczka, M. 1999, ApJ, 524, 179

\reference{} Decourchelle, A., Sauvageot, J. L., Ballet, J., \&
Aschenbach, B., 1997, A\&A, 326, 811

\reference{} Desai, P., et al, 2005, ApJ, 625, L59

\reference{} Ghavamian, P., Raymond, J. C., Smith, R. C., \& Hartigan,
P., 2001, ApJ, 547, 995

\reference{} Graham, J.R., Levenson, N.A., Hester, J.J., Raymond, J.C. \& Petre, R. 1995, ApJ, 444, 787

\reference{} Fesen, R.A., \& Itoh, H. 1985, ApJ, 293, 43

\reference{} Fesen, R.A., Kwitter, K.B. \& Downes, R.A. 1992, AJ, 104, 719
 
\reference{} Hester, J.J., Raymond, J.C. \& Blair, W.P. 1994, ApJ, 420, 721

\reference{} Katsuda, S., Tsunemi, H., Kimura, M., \& Mori, K.,
2008, ApJ, 680, 1198

\reference{} Levenson, N. A., et al., 1997, ApJ, 484, 304

\reference{} Levenson, N. A., Graham, J. R., Keller, L. D., \& Richter,
M. J, 1998, 118, 541

\reference{} Levenson, N. A., Graham, J. R., \& Snowden, S. L., 1999,
ApJ, 526, 874

\reference{} Levenson, N. A., Graham, J. R., \& Walters, J. L., 2002,
ApJ, 576, 798 

\reference{} Long, K.S., Blair, W.P., Vancura, O., Bowers, C., Davidsen, A.F.
\& Raymond, J.C. 1992, ApJ, 400, 214  

\reference{} Malkov, M. A., Diamond, P. H., \& V$\ddot{o}$lk, H. J., 2000,
ApJ, 533, L171

\reference{} McEntaffer, R. L., \& Cash, W., 2008, arXiv.0801.4552v1

\reference{} McKee, C. F., \& Hollenbach, D. J., 1980,
Ann. Rev. A\&A, 18, 219

\reference{} Minkowski, R. 1958, Rev. Mod. Phys., 30, 1048

\reference{} Mitaya, E., Katsuda, S., Tsunemi, H., Hughes, J. P.,
Kokubun, M., \& Porter, F. S., 2007, PASJ, 59, S163

\reference{} Nemes, N., Tsunemi, H., \$ Mityata, E., 2008, ApJ, 675, 1293

\reference{} Prieto, M. A., Hasinger, G., \& Snowden, S. L., 1996,
Astron. Astrophys. Suppl. Ser. 120, 187-193

\reference{} Raymond, J.C., Davis, M., Gull, T.R. \& Parker, R.A.R. 1980, 
ApJL, 238, L21

\reference{} Raymond, J. C, Ghavamian, P., Sankrit, R., Blair, W. P.,
\& Curiel, S., 2003, ApJ, 584, 770

\reference{} Raymond, J. C., 2008.n {\it Particle Acceleration and Transport in the Heliosphere and Beyondi, AIP Conference Proceedings 1039}, G. Li, Q. Hu, O. Verkhoglyadova, G.P. Zank, R.P. Lin and J.  Luhmann, eds. (Melville, NY: AIP), p. 433

\reference{} Raymond, J. C, \& Smith, B. W., 1977, ApJS, 35, 419

\reference{} Sakhibov, F.H. \& Smirnov, M.A. 1983, Sov. Astr., 27, 395

\reference{} Shelton, R.L., Cox, D.P., Maciejewski, W., Smith, R.K.,
Plewa, T., Pawl, A. \& R\'{o}\.{z}yczka, M. 1999, ApJ, 524, 192

\reference{} Shull, P.J. \& Hippelein, H.H. 1991, ApJ, 383, 714

\reference{} Slavin, J.D. \& Cox, D.P. 1992, ApJ, 392, 131

\reference{} Smith, R. K., Brickhouse, N. S., Liedahl, D. A., \& Raymond, J. C. 2001, ApJ, 556, L91

\reference{} Treffers, R.R. 1981, ApJ 250, 213

\reference{} Tsunemi, H., Katsuda, S., Nemes, N., \& Miller, E. D.,
2007, ApJ, 671, 1717

\reference{} Turner, T., J., 1996, OGIP Memo OGIP/94-010

\reference{} van Adelsberg, M., Heng, K., McCray, R. \& Raymond, J.C. 2008, 
ApJ, in press

\reference{} Vedder, P. W., Canizares, C. R., Markert, T. H., \&
Pradhan, A. K., 1986, ApJ, 307, 269

\end{references}
\end{document}